\documentclass[aps,prc,preprint,groupedaddress,showpacs,floatfix]{revtex4}
\usepackage{graphicx}
\usepackage{amsmath,amssymb,bm,ascmac,calc}

\begin{document}


\title{Semi-realistic nucleon-nucleon interactions
with improved neutron-matter properties}


\author{H. Nakada}
\email[E-mail:\,\,]{nakada@faculty.chiba-u.jp}
\affiliation{Department of Physics, Graduate School of Science,
 Chiba University\\
Yayoi-cho 1-33, Inage, Chiba 263-8522, Japan}


\date{\today}

\begin{abstract}
New parameter-sets of the semi-realistic nucleon-nucleon interaction
are developed,
by modifying the M3Y interaction but maintaining the tensor channels
and the longest-range central channels.
The modification is made so as to reproduce
microscopic results of neutron-matter energies,
in addition to the measured binding energies of doubly magic nuclei
including $^{100}$Sn
and the even-odd mass differences of the $Z=50$ and $N=82$ nuclei
in the self-consistent mean-field calculations.
Separation energies of the proton- or neutron-magic nuclei
are shown to be in fair agreement with the experimental data.
With the new parameter-sets M3Y-P6 and P7,
the isotropic spin-saturated symmetric nuclear matter remains stable
in the density range as wide as $\rho\lesssim 6\rho_0$,
while keeping desirable results of the previous parameter-set
on finite nuclei.
Isotope shifts of the Pb nuclei
and tensor-force effects on shell structure are discussed.
\end{abstract}

\pacs{21.30.Fe, 21.60.Jz, 21.65.-f, 21.10.Dr}

\maketitle



\section{Introduction\label{sec:intro}}

As exotic natures of unstable nuclei such as the new magic numbers
and the neutron halos are disclosed by experiments,
microscopic studies based on the nucleon-nucleon ($NN$) interaction
become even more desired in nuclear structure physics.
While the fully microscopic $NN$ (and $NNN$) interaction
is still too complicated
to cover large volume of nuclei in the periodic table
despite significant progress~\cite{ref:micro,ref:NCSM,ref:HPDH08},
the semi-realistic $NN$ interactions have been developed~\cite{ref:Nak03,
ref:Nak08b} by modifying
the Michigan 3-range Yukawa (M3Y) interaction~\cite{ref:M3Y,ref:M3Y-P},
which was originated from Brueckner's $G$-matrix at the nuclear surface
and expressed by the Yukawa functions.
The modification has been made so that
the saturation and the spin-orbit ($\ell s$) splitting should be
reproduced within the mean-field approximation (MFA).
Owing to the recently developed numerical methods~\cite{ref:NS02,
ref:Nak06,ref:Nak08,ref:NMYM09},
self-consistent calculations in the MFA~\cite{ref:Nak04,ref:Nak09,ref:Nak10b}
and in the random-phase approximation (RPA)~\cite{ref:Shi08}
have been implemented using the semi-realistic interactions.
Among the advantages of the M3Y-type semi-realistic interactions
is that they contain realistic tensor channels
as well as correct longest-range central channels
originating from the one-pion exchange,
which have been pointed out
to play significant roles in $Z$- or $N$-dependence
of the shell structure~\cite{ref:Vst,ref:Vtn}.
The semi-realistic interactions are suitable to investigate
effects of these channels within the self-consistent MFA
and RPA~\cite{ref:Nak08b}.

While the parameter-sets of the M3Y-type interactions
in Refs.~\cite{ref:Nak03,ref:Nak08b,ref:Nak10}
were adjusted to the data on the nuclear structure,
some of them have been applied to the nuclear reactions~\cite{ref:Khoa-pv}
and to the neutron stars~\cite{ref:LTKM11} as well.
In studying structure of the neutron stars,
density-dependence of the symmetry energy
is crucially important~\cite{ref:LP07}.
It has been pointed out
that the symmetry energy at low density is significant in nuclear reactions:
\textit{e.g.} the charge-exchange reactions~\cite{ref:Khoa-pv}
and the multi-fragmentation processes~\cite{ref:Ono03}.
The symmetry energy at low density may also affect
the so-called pygmy dipole resonance in neutron-rich nuclei~\cite{ref:PDR-L}.
However, the symmetry energy, particularly its density-dependence,
was not sufficiently reliable in the previous parameter-sets
in Refs.~\cite{ref:Nak08b,ref:Nak10}
as pointed out in Ref.~\cite{ref:TKG09},
giving rise to instability of the symmetric nuclear matter
at the density $\rho\gtrsim 0.6\,\mathrm{fm}^{-3}$,
which is not consistent with microscopic calculations~\cite{ref:APR98}.
In this article, we shall propose new parameter-sets
of the M3Y-type semi-realistic $NN$ interaction.
As far as the energy of the symmetric nuclear matter is fixed,
the symmetry energy at each density is well connected
to the energy of the neutron matter.
The new parameters are determined by fitting the neutron-matter energy
to microscopic result in Ref.~\cite{ref:FP81} (FP)
or \cite{ref:APR98} (APR).
Moreover, we additionally take into consideration
the binding energy of $^{100}$Sn.
As argued later, the symmetry energy at the saturation point
tends to be fixed with good precision by fitting the parameters
both to $^{100}$Sn and $^{132}$Sn.
The symmetry energy is thus constrained
to good degree in the new parameter-sets.
Corresponding to the microscopic results on the neutron-matter energy,
we obtain two parameter-sets
`M3Y-P6' (fitted to FP) and `M3Y-P7' (to APR).
Although the parameters are determined from a limited number of data,
they will be useful for investigating various aspects of nuclear properties,
as will be illustrated by separation energies
of the proton- and neutron-magic nuclei
and by $Z$- or $N$-dependence of shell structure.

\section{M3Y-type interaction\label{sec:M3Y}}

We take a non-relativistic isoscalar nuclear Hamiltonian of
\begin{equation}
H_N = K + V_N\,;\quad K = \sum_i \frac{\mathbf{p}_i^2}{2M}\,,\quad
V_N = \sum_{i<j} v_{ij}\,,
\label{eq:H_N}\end{equation}
with $i$ and $j$ representing the indices of individual nucleons.
We set $M=(M_p+M_n)/2$ throughout this paper,
where $M_p$ ($M_n$) is the mass of a proton (a neutron)~\cite{ref:PDG06}.
For the effective $NN$ interaction $v_{ij}$,
the following form is considered,
\begin{eqnarray} v_{ij} &=& v_{ij}^{(\mathrm{C})}
 + v_{ij}^{(\mathrm{LS})} + v_{ij}^{(\mathrm{TN})}
 + v_{ij}^{(\mathrm{DD})}\,;\nonumber\\
v_{ij}^{(\mathrm{C})} &=& \sum_n \big(t_n^{(\mathrm{SE})} P_\mathrm{SE}
+ t_n^{(\mathrm{TE})} P_\mathrm{TE} + t_n^{(\mathrm{SO})} P_\mathrm{SO}
+ t_n^{(\mathrm{TO})} P_\mathrm{TO}\big)
 f_n^{(\mathrm{C})} (r_{ij})\,,\nonumber\\
v_{ij}^{(\mathrm{LS})} &=& \sum_n \big(t_n^{(\mathrm{LSE})} P_\mathrm{TE}
 + t_n^{(\mathrm{LSO})} P_\mathrm{TO}\big)
 f_n^{(\mathrm{LS})} (r_{ij})\,\mathbf{L}_{ij}\cdot
(\mathbf{s}_i+\mathbf{s}_j)\,,\nonumber\\
v_{ij}^{(\mathrm{TN})} &=& \sum_n \big(t_n^{(\mathrm{TNE})} P_\mathrm{TE}
 + t_n^{(\mathrm{TNO})} P_\mathrm{TO}\big)
 f_n^{(\mathrm{TN})} (r_{ij})\, r_{ij}^2 S_{ij}\,,\nonumber\\
v_{ij}^{(\mathrm{DD})} &=& \big(t_\rho^{(\mathrm{SE})} P_\mathrm{SE}\cdot
 [\rho(\mathbf{r}_i)]^{\alpha^{(\mathrm{SE})}}
 + t_\rho^{(\mathrm{TE})} P_\mathrm{TE}\cdot
 [\rho(\mathbf{r}_i)]^{\alpha^{(\mathrm{TE})}}\big)
 \,\delta(\mathbf{r}_{ij})\,,
\label{eq:effint}\end{eqnarray}
where $\mathbf{s}_i$ is the spin operator of the $i$-th nucleon,
$\mathbf{r}_{ij}= \mathbf{r}_i - \mathbf{r}_j$,
$r_{ij}=|\mathbf{r}_{ij}|$,
$\mathbf{p}_{ij}= (\mathbf{p}_i - \mathbf{p}_j)/2$,
$\mathbf{L}_{ij}= \mathbf{r}_{ij}\times \mathbf{p}_{ij}$,
and $\rho(\mathbf{r})$ denotes the nucleon density.
The tensor operator is defined by
$S_{ij}= 4\,[3(\mathbf{s}_i\cdot\hat{\mathbf{r}}_{ij})
(\mathbf{s}_j\cdot\hat{\mathbf{r}}_{ij})
- \mathbf{s}_i\cdot\mathbf{s}_j ]$
with $\hat{\mathbf{r}}_{ij}=\mathbf{r}_{ij}/r_{ij}$.
The projection operators
on the singlet-even (SE), triplet-even (TE), singlet-odd (SO)
and triplet-odd (TO) two-particle states are
\begin{eqnarray}
P_\mathrm{SE} = \frac{1-P_\sigma}{2}\,\frac{1+P_\tau}{2}\,,
\quad P_\mathrm{TE} = \frac{1+P_\sigma}{2}\,\frac{1-P_\tau}{2}\,,
\nonumber\\
P_\mathrm{SO} = \frac{1-P_\sigma}{2}\,\frac{1-P_\tau}{2}\,.
\quad P_\mathrm{TO} = \frac{1+P_\sigma}{2}\,\frac{1+P_\tau}{2}\,,
\label{eq:proj_T}\end{eqnarray}
where $P_\sigma$ ($P_\tau$) expresses
the spin (isospin) exchange operator.
The Yukawa function $f_n(r)=e^{-\mu_n r}/\mu_n r$
is assumed for all channels except $v^{(\mathrm{DD})}$.
The density-dependent contact term $v^{(\mathrm{DD})}$ is added
in order to reproduce the saturation properties.
Physically, $v^{(\mathrm{DD})}$ may carry
effects of the $NNN$ interaction
and of the density-dependence that is dropped in the original M3Y interaction.

We start from the M3Y-Paris interaction~\cite{ref:M3Y-P},
which will be denoted by M3Y-P0 in this article
as in Ref.~\cite{ref:Nak03}.
The range parameters $\mu_n$ of M3Y-P0 are maintained
in any of $v^{(\mathrm{C})}$, $v^{(\mathrm{LS})}$ and $v^{(\mathrm{TN})}$.
As in M3Y-P0, the longest-range part in $v^{(\mathrm{C})}$
is kept identical to the central channels
of the one-pion exchange potential (OPEP), $v^{(\mathrm{C})}_\mathrm{OPEP}$.
Although the $\ell s$ splitting plays
a significant role in the nuclear shell structure,
the $G$-matrix is known to underestimate the $\ell s$ splitting.
Even though effects beyond the MFA may cure this problem~\cite{ref:LS},
we introduce an overall enhancement factor to $v^{(\mathrm{LS})}$
in order to describe the shell structure within the MFA.
Effects of the tensor force on the single-particle (s.p.) levels
could be relevant to the new magic numbers
in unstable nuclei~\cite{ref:Nak08b,ref:Vtn}.
We keep $v^{(\mathrm{TN})}$ without any modification from M3Y-P0.
Because of this $v^{(\mathrm{TN})}$ having realistic nature,
the present M3Y-type interactions are useful to investigate
the tensor-force effects within the MFA and RPA,
as shown in Refs.~\cite{ref:Nak09,ref:Nak10b,ref:Shi08}
with the previous parameter-sets.
The parameters in M3Y-P6 and P7 are tabulated
in Table~\ref{tab:param_M3Y}, together with M3Y-P0.

\begin{table}
~\vspace*{-2.3cm}
\begin{center}
\caption{Parameters of M3Y-type interactions.
\label{tab:param_M3Y}}
\begin{tabular}{ccr@{.}lr@{.}lr@{.}lr@{.}l}
\hline\hline
parameters && \multicolumn{2}{c}{~~M3Y-P0~~} &
 \multicolumn{2}{c}{~~M3Y-P6~~} & \multicolumn{2}{c}{~~M3Y-P7~~} \\
 \hline
$1/\mu_1^{(\mathrm{C})}$ &(fm)& $0$&$25$ & $0$&$25$ & $0$&$25$ \\
$t_1^{(\mathrm{SE})}$ &(MeV)& $11466$& & $10766$& & $10655$& \\
$t_1^{(\mathrm{TE})}$ &(MeV)& $13967$& & $8474$& & $9592$& \\
$t_1^{(\mathrm{SO})}$ &(MeV)& $-1418$& & $-728$& & $11510$& \\
$t_1^{(\mathrm{TO})}$ &(MeV)& $11345$& & $12453$& & $13507$& \\
$1/\mu_2^{(\mathrm{C})}$ &(fm)& $0$&$40$ & $0$&$40$ & $0$&$40$ \\
$t_2^{(\mathrm{SE})}$ &(MeV)& $-3556$& & $-3520$& & $-3556$& \\
$t_2^{(\mathrm{TE})}$ &(MeV)& $-4594$& & $-4594$& & $-4594$& \\
$t_2^{(\mathrm{SO})}$ &(MeV)& $950$& & $1386$& & $1283$& \\
$t_2^{(\mathrm{TO})}$ &(MeV)& $-1900$& & $-1588$& & $-1812$& \\
$1/\mu_3^{(\mathrm{C})}$ &(fm)& $1$&$414$ & $1$&$414$ & $1$&$414$ \\
$t_3^{(\mathrm{SE})}$ &(MeV)& $-10$&$463$ & $-10$&$463$ & $-10$&$463$ \\
$t_3^{(\mathrm{TE})}$ &(MeV)& $-10$&$463$ & $-10$&$463$ & $-10$&$463$ \\
$t_3^{(\mathrm{SO})}$ &(MeV)& $31$&$389$ & $31$&$389$ & $31$&$389$ \\
$t_3^{(\mathrm{TO})}$ &(MeV)& $3$&$488$ & $3$&$488$ & $3$&$488$ \\
$1/\mu_1^{(\mathrm{LS})}$ &(fm)& $0$&$25$ & $0$&$25$ & $0$&$25$ \\
$t_1^{(\mathrm{LSE})}$ &(MeV)& $-5101$& & $-11222$&$2$ & $-11732$&$3$ \\
$t_1^{(\mathrm{LSO})}$ &(MeV)& $-1897$& & $-4173$&$4$ & $-4363$&$1$ \\
$1/\mu_2^{(\mathrm{LS})}$ &(fm)& $0$&$40$ & $0$&$40$ & $0$&$40$ \\
$t_2^{(\mathrm{LSE})}$ &(MeV)& $-337$& & $-741$&$4$ & $-775$&$1$ \\
$t_2^{(\mathrm{LSO})}$ &(MeV)& $-632$& & $-1390$&$4$ & $-1453$&$6$ \\
$1/\mu_1^{(\mathrm{TN})}$ &(fm)& $0$&$40$ & $0$&$40$ & $0$&$40$ \\
$t_1^{(\mathrm{TNE})}$ &(MeV$\cdot$fm$^{-2}$)& $-1096$& & $-1096$& & $-1096$& \\
$t_1^{(\mathrm{TNO})}$ &(MeV$\cdot$fm$^{-2}$)& $244$& & $244$& & $244$& \\
$1/\mu_2^{(\mathrm{TN})}$ &(fm)& $0$&$70$ & $0$&$70$ & $0$&$70$ \\
$t_2^{(\mathrm{TNE})}$ &(MeV$\cdot$fm$^{-2}$)& $-30$&$9$ & $-30$&$9$ & $-30$&$9$ \\
$t_2^{(\mathrm{TNO})}$ &(MeV$\cdot$fm$^{-2}$)& $15$&$6$ & $15$&$6$ & $15$&$6$ \\
$\alpha^{(\mathrm{SE})}$ && \multicolumn{2}{c}{---} &
 \multicolumn{2}{c}{$1$} & \multicolumn{2}{c}{$1$} \\
$t_\rho^{(\mathrm{SE})}$ &(MeV$\cdot$fm$^3$)& $0$& & $384$& & $830$& \\
$\alpha^{(\mathrm{TE})}$ && \multicolumn{2}{c}{---} &
 \multicolumn{2}{c}{$1/3$} & \multicolumn{2}{c}{$1/3$} \\
$t_\rho^{(\mathrm{TE})}$ &(MeV$\cdot$fm)& $0$& & $1930$& & $1478$& \\
\hline\hline
\end{tabular}
\end{center}
\end{table}

\section{Properties of nuclear matter\label{sec:NMprop}}

We first apply the new semi-realistic interactions
to the infinite nuclear matter in the Hartree-Fock (HF) approximation.
Notice that only $v^{(\mathrm{C})}+v^{(\mathrm{DD})}$
in Eq. (\ref{eq:effint}) contributes to the nuclear matter properties
in the MFA.
Energy of the nuclear matter is a function of the following variables,
\begin{eqnarray}
 \rho &=& {\displaystyle\sum_{\sigma\tau}} \rho_{\tau\sigma}\,,
  \nonumber\\
 \eta_s &=& \frac{{\displaystyle\sum_{\sigma\tau}}
  \sigma\rho_{\tau\sigma}}{\rho}
  ~=~ \frac{\rho_{p\uparrow}-\rho_{p\downarrow}+\rho_{n\uparrow}
  -\rho_{n\downarrow}}{\rho}\,, \nonumber\\
 \eta_t &=& \frac{{\displaystyle\sum_{\sigma\tau}}
  \tau\rho_{\tau\sigma}}{\rho}
  ~=~ \frac{\rho_{p\uparrow}+\rho_{p\downarrow}-\rho_{n\uparrow}
  -\rho_{n\downarrow}}{\rho}\,, \nonumber\\
 \eta_{st} &=& \frac{{\displaystyle\sum_{\sigma\tau}}
  \sigma\tau\rho_{\tau\sigma}}{\rho}
  ~=~ \frac{\rho_{p\uparrow}-\rho_{p\downarrow}-\rho_{n\uparrow}
  +\rho_{n\downarrow}}{\rho}\,,
\end{eqnarray}
where $\tau=p,n$ and $\sigma=\uparrow,\downarrow$
are sometimes substituted by $\pm 1$ without confusion.
As we restrict ourselves to the properties at zero temperature,
the density depending on the spin and the isospin $\rho_{\tau\sigma}$ 
is related to the Fermi momentum $k_{\mathrm{F}\tau\sigma}$ via
\begin{equation}
 \rho_{\tau\sigma} = \frac{1}{6\pi^2} k_{\mathrm{F}\tau\sigma}^3\,.
\end{equation}
Basic formulas to calculate the nuclear matter energy and its derivatives
for given $k_{\mathrm{F}\tau\sigma}$ were derived
in Ref.~\cite{ref:Nak03}.
Note that the superfluidity barely influences the nuclear matter energy,
even if it takes place.

We shall denote energy per nucleon ($E/A$) by $\mathcal{E}$,
where $E$ is the expectation value of $H_N$ for the nuclear matter.
The spin-saturated symmetric matter is
characterized by $\eta_s=\eta_t=\eta_{st}=0$,
for which we represent $k_{\mathrm{F}\tau\sigma}$ simply by $k_\mathrm{F}$.
Minimization of $\mathcal{E}(\rho)$,
\begin{equation}
 \left.\frac{\partial\mathcal{E}}{\partial\rho}\right\vert_0
 =\left.\frac{\partial\mathcal{E}}{\partial k_\mathrm{F}}\right\vert_0
 =0\,,
\end{equation}
determines the saturation density $\rho_0$
(equivalently, $k_{\mathrm{F}0}$) and the saturation energy $\mathcal{E}_0$.
The expression $~\vert_0$ indicates evaluation at the saturation point.

\begin{figure}
\includegraphics[scale=0.9]{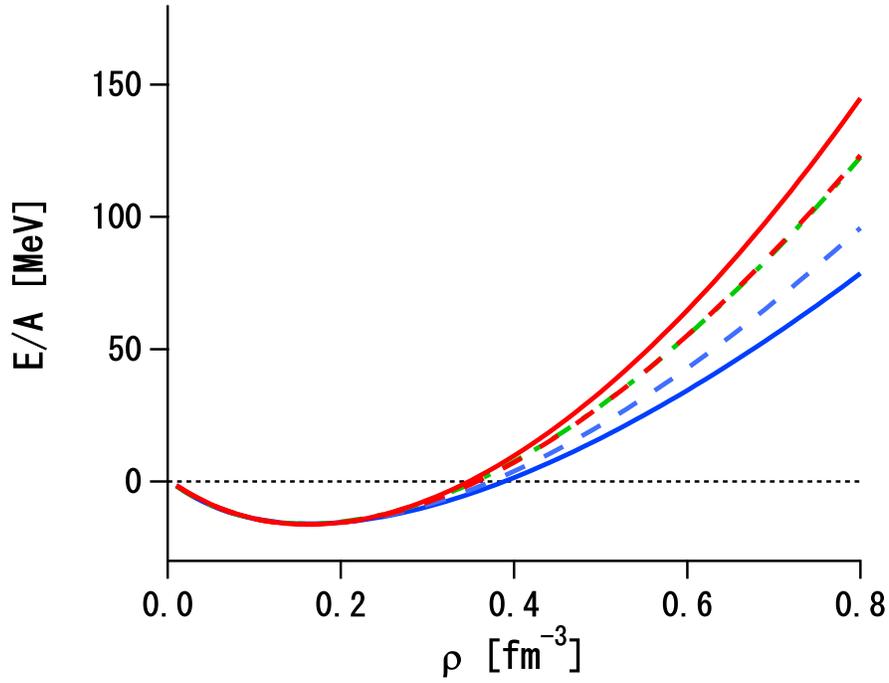}
\caption{(Color online) $\mathcal{E}=E/A$ \textit{vs.} $\rho$
in the symmetric nuclear matter,
calculated with M3Y-P6 (red dashed line), M3Y-P7 (red solid line),
D1S (blue solid line), D1M (blue dashed line)
and SLy5 (green dot-dashed line).
\label{fig:NME_M3Ya}}
\end{figure}

We depict $\mathcal{E}(\rho)$
for the spin-saturated symmetric nuclear matter in Fig.~\ref{fig:NME_M3Ya},
up to $\rho\approx 5\rho_0$.
The results of the new semi-realistic (\textit{i.e.} M3Y-P6 and P7)
interactions are compared with those of the Skyrme SLy5~\cite{ref:SLy},
the Gogny D1S~\cite{ref:D1S} and D1M~\cite{ref:D1M} interactions.
While all effective interactions give close $\mathcal{E}(\rho)$
at $\rho\lesssim \rho_0$,
interaction-dependence is visible at $\rho\gtrsim 2\rho_0$,
though the SLy5 energy almost coincides with the M3Y-P6 one.

\begin{figure}
\includegraphics[scale=0.9]{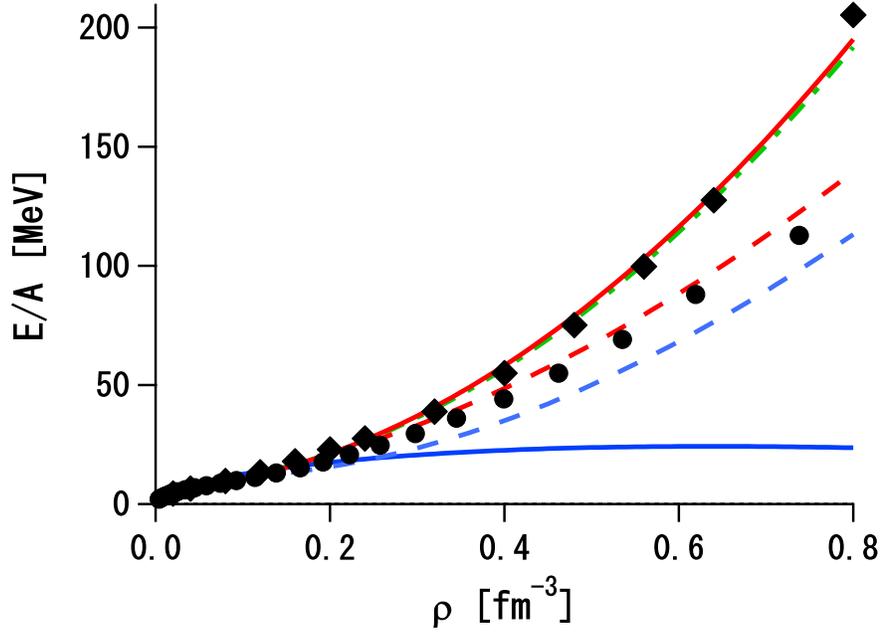}
\caption{(Color online) $\mathcal{E}=E/A$ \textit{vs.} $\rho$
in the pure neutron matter.
Circles and diamonds represent the FP and APR results, respectively.
See Fig.~\protect\ref{fig:NME_M3Ya} for the other conventions.
\label{fig:NME_M3Yc}}
\end{figure}

Energy per nucleon in the spin-saturated neutron matter
(\textit{i.e.} $\eta_t=-1$, $\eta_s=\eta_{st}=0$) is shown
in Fig.~\ref{fig:NME_M3Yc}.
The FP~\cite{ref:FP81} and APR~\cite{ref:APR98} results,
to which M3Y-P6 and P7 are respectively fitted, are also presented.
Having been fitted to a microscopic result~\cite{ref:WFF88} as well,
the energy with SLy5 is close to that with M3Y-P7 at any $\rho$.
The stronger $\rho$ dependence in M3Y-P6 and P7
than in D1S and D1M
originates from the choice $\alpha^{(\mathrm{SE})}=1$ in $v^{(\mathrm{DD})}$,
and enables us to reproduce the microscopic results.
Since $v^{(\mathrm{DD})}$ drives repulsion in the SE channel at high $\rho$
but does not in the TO channel,
the interactions having the form of Eq.~(\ref{eq:effint}) may give rise to
the spin-polarized phase at high $\rho$ in the pure neutron matter.
However, the transition to the spin-polarized phase is delayed
until $\rho\approx 9\rho_0$ ($20\rho_0$) for M3Y-P7 (P6),
almost irrelevant even to the neutron stars.

The symmetry energy is defined by the second derivative of $\mathcal{E}$
with respect to $\eta_t$ for the spin-saturated matter,
\begin{equation}
 a_t(\rho) = \left. \frac{1}{2} \frac{\partial^2\mathcal{E}}{\partial\eta_t^2}
	\right\vert_\rho\,.
\end{equation}
Here $~\vert_\rho$ indicates $\eta_t=\eta_s=\eta_{st}=0$
but $\rho$ left as a variable.
The symmetry energy at the saturation point $a_t(\rho_0)$
is denoted by $a_{t0}$.
Analogously, we define
\begin{equation}
 a_s(\rho) = \left. \frac{1}{2} \frac{\partial^2\mathcal{E}}{\partial\eta_s^2}
	\right\vert_\rho\,, \quad
 a_{st}(\rho) = \left. \frac{1}{2} \frac{\partial^2\mathcal{E}}
 {\partial\eta_{st}^2}\right\vert_\rho\,,
\end{equation}
and $a_{s0}=a_s(\rho_0)$, $a_{st0}=a_{st}(\rho_0)$.
The incompressibility at the saturation point is obtained by
\begin{equation}
 \mathcal{K}_0 = k_\mathrm{F}^2 \left.\frac{\partial^2\mathcal{E}}
  {\partial k_\mathrm{F}^2}\right\vert_0
 = 9\rho^2 \left.\frac{\partial^2\mathcal{E}}{\partial\rho^2}
       \right\vert_0\,.
\end{equation}
The effective mass ($k$-mass) is defined by a derivative of
the s.p. energy $\varepsilon(\mathbf{k}\sigma\tau)$.
We denote the effective mass at the saturation point by $M^\ast_0$,
which is given as
\begin{equation}
 \left.\frac{\partial\varepsilon(\mathbf{k}\sigma\tau)}{\partial k}
 \right\vert_0 = \frac{k_{\mathrm{F}0}}{M^\ast_0}\,.
\label{eq:M*}\end{equation}
These characteristic quantities
calculated from the new semi-realistic interactions
are tabulated in Tables~\ref{tab:NMsat}.
Those from D1S, D1M and SLy5 are also displayed for comparison.
Also compare them with the empirical values
$k_\mathrm{F0}\sim 1.33-1.34\,\mathrm{fm}^{-1}$,
$\mathcal{E}_0\sim -16\,\mathrm{MeV}$,
$\mathcal{K}\sim 240\,\mathrm{MeV}$~\cite{ref:SKC06}
and $a_{t0}\sim 30\,\mathrm{MeV}$~\cite{ref:Dan03}.
It is remarked that the fit both to $^{100}$Sn and $^{132}$Sn
(see Table~\ref{tab:DMprop})
well constrains $a_{t0}$ in the M3Y-type interactions with good precision,
to $a_{t0}\approx 32\,\mathrm{MeV}$.
This $a_{t0}$ value is in harmony with $a_{t0}$ of SLy5,
though not so with $a_{t0}$ of D1M and of a recently proposed
Skyrme-type energy density functional UNDEF1~\cite{ref:UNDEF1}.
The effective mass ($M_0^\ast\approx 0.6\,M$)
of the current M3Y-type interactions
is not much changeable,
which does not contradict to a microscopic result~\cite{ref:Mahaux}
but is lower than the value
that reproduces collective excitations in the RPA
(\textit{e.g.} the D1M value).

\begin{table}
\begin{center}
\caption{Nuclear matter properties at the saturation point.
\label{tab:NMsat}}
\begin{tabular}{ccrrrrr}
\hline\hline
&&~~M3Y-P6 &~~M3Y-P7 &~~~~D1S~~ &~~~~D1M~~ &~~~~SLy5~ \\ \hline
$k_{\mathrm{F}0}$ & (fm$^{-1}$) & $1.340$~~& $1.340$~~
 & $1.342$~~& $1.346$~~& $1.334$~~\\
$\mathcal{E}_0$ & (MeV) & $-16.24$~~& $-16.22$~~
 & $-16.01$~~& $-16.02$~~& $-15.98$~~\\
$\mathcal{K}_0$ & (MeV) & $239.7$~~& $254.7$~~
 & $202.9$~~& $225.0$~~& $229.9$~~\\
$M^\ast_0/M$ && $0.596$~~& $0.589$~~
 & $0.697$~~& $0.746$~~& $0.697$~~\\
$a_{t0}$ & (MeV) & $32.14$~~& $31.74$~~
 & $31.12$~~& $28.55$~~& $32.03$~~\\
$a_{s0}$ & (MeV) & $26.47$~~& $23.04$~~
 & $26.18$~~& $16.56$~~& $37.47$~~\\
$a_{st0}$ & (MeV) & $41.00$~~& $43.30$~~
 & $29.13$~~& $28.71$~~& $15.15$~~\\
$\mathcal{Q}_{0}$ & (MeV) & $-378.0$~~& $-320.1$~~
 & $-515.7$~~& $-459.0$~~& $-363.9$~~\\
$\mathcal{L}_{t0}$ & (MeV) & $44.64$~~& $51.53$~~
 & $22.44$~~& $24.83$~~& $48.27$~~\\
\hline\hline
\end{tabular}
\end{center}
\end{table}

As mentioned in Introduction,
$\rho$-dependence of the symmetry energy attracts interest.
The first derivative of $a_t(\rho)$ at $\rho_0$ is under debate,
which is parametrized as
\begin{equation} \mathcal{L}_{t0} = 3\left.\frac{d}{d\rho} a_t(\rho)\right\vert_0
 = \left.\frac{1}{2}k_\mathrm{F}\frac{\partial^3\mathcal{E}}
  {\partial k_\mathrm{F}\,\partial\eta_t^2}\right\vert_0
 = \left.\frac{3}{2}\rho\frac{\partial^3\mathcal{E}}
  {\partial\rho\,\partial\eta_t^2}\right\vert_0\,.
\end{equation}
The characteristic coefficient $\mathcal{L}_{t0}$,
along with the third derivative of $\mathcal{E}$ with respect to $\rho$
that is denoted by $\mathcal{Q}_0$,
\begin{equation} \mathcal{Q}_0 = \left.k_\mathrm{F}^3
  \frac{\partial^3\mathcal{E}}{\partial k_\mathrm{F}^3}\right\vert_0
 = \left.27\rho^3\frac{\partial^3\mathcal{E}}{\partial\rho^3}\right\vert_0\,,
\end{equation}
are also presented in Table~\ref{tab:NMsat}.
It is noteworthy that M3Y-P6 and P7 have higher $\mathcal{L}_{t0}$
than D1S and D1M,
in contrast to the previous parameter-set M3Y-P5~\cite{ref:Nak08b}
that has comparable $\mathcal{L}_{t0}$ to D1S and D1M.
The higher $\mathcal{L}_{t0}$ values seem favorable
for describing the low-lying $E1$ strengths~\cite{ref:PDR-L}.

\begin{figure}
\includegraphics[scale=0.7]{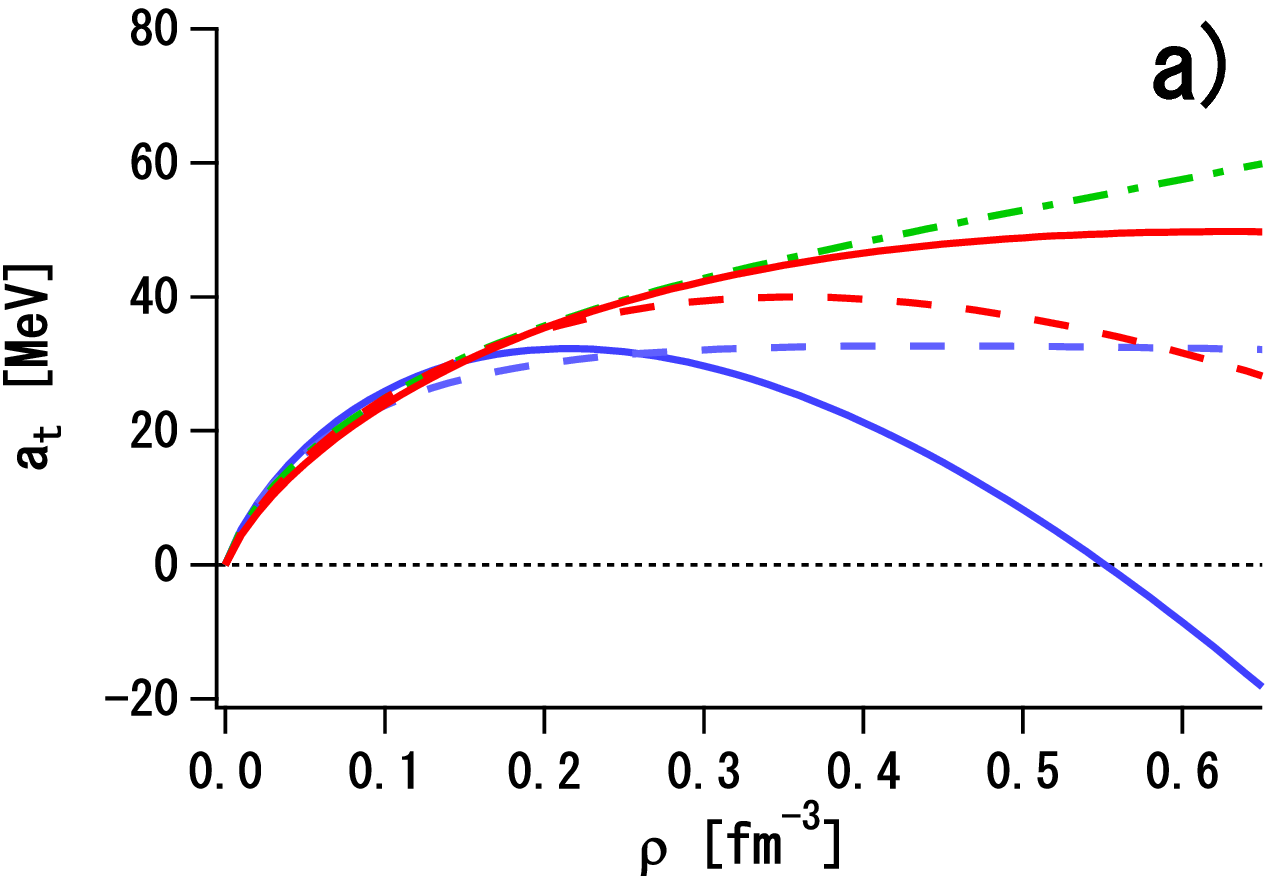}\\
\includegraphics[scale=0.7]{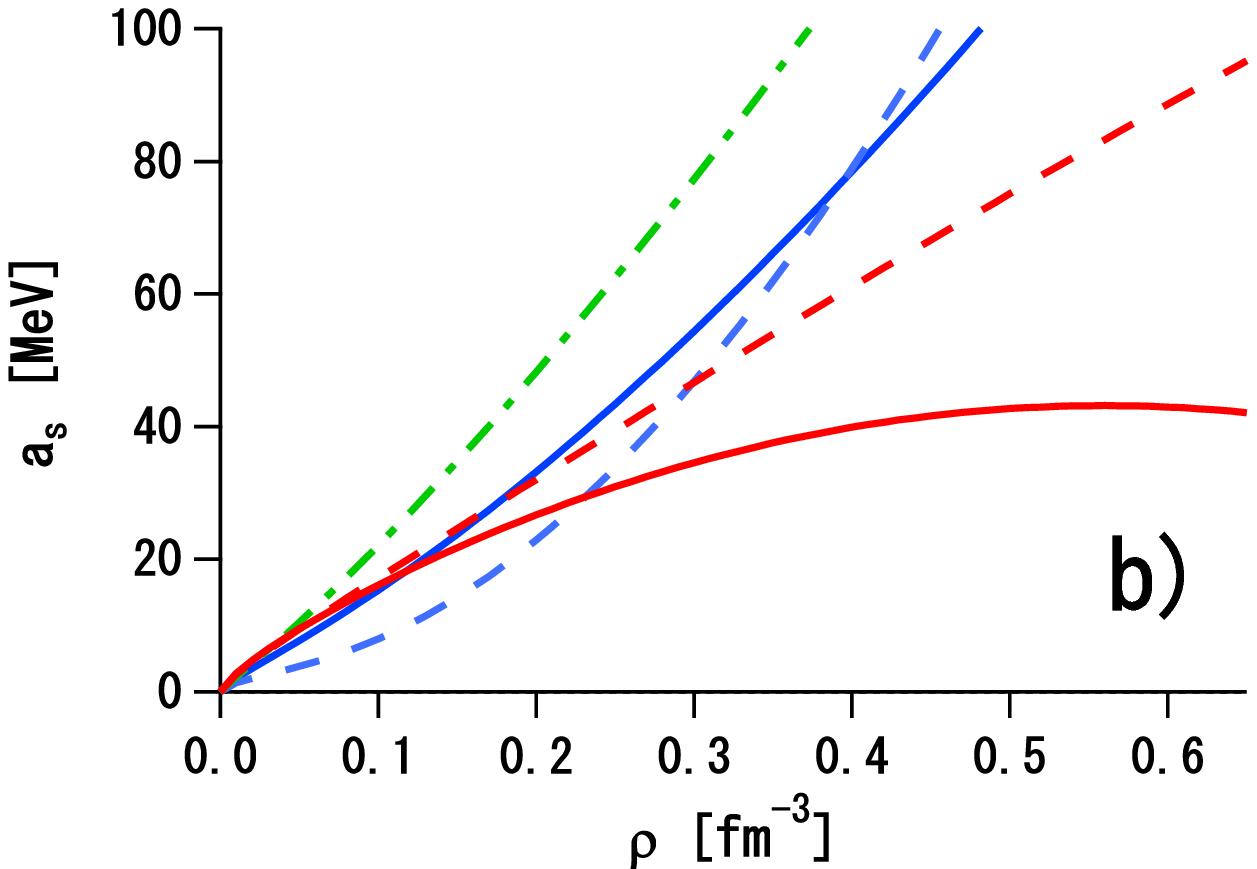}\\
\includegraphics[scale=0.7]{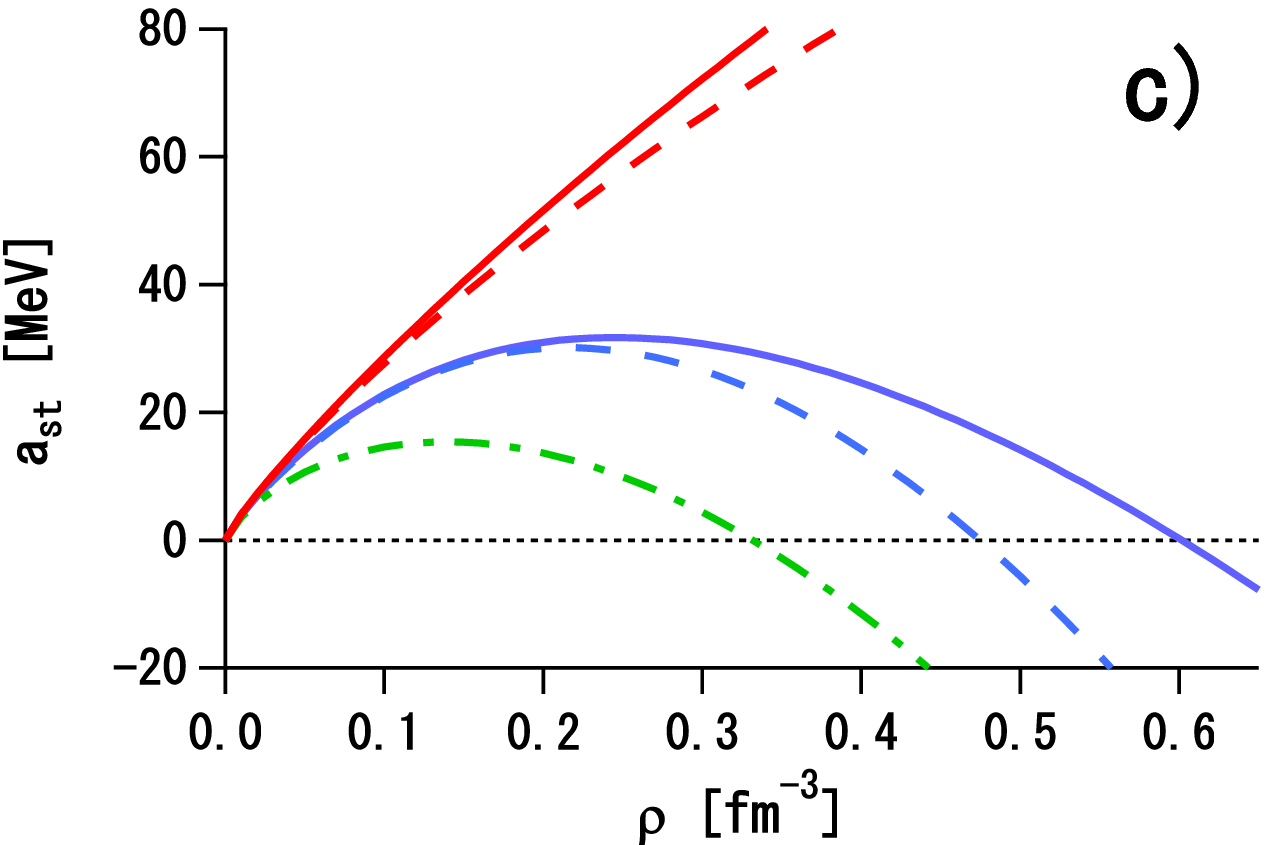}
\caption{(Color online) a) $a_t(\rho)$, b) $a_s(\rho)$ and c) $a_{st}(\rho)$
in the symmetric nuclear matter.
See Fig.~\protect\ref{fig:NME_M3Ya} for the conventions.
\label{fig:a_st}}
\end{figure}

The symmetry energy $a_t(\rho)$ in a wider region of $\rho$
is depicted in Fig.~\ref{fig:a_st},
along with $a_s(\rho)$ and $a_{st}(\rho)$.
If any of $a_t(\rho)$, $a_s(\rho)$ or $a_{st}(\rho)$ is negative,
the spin-saturated symmetric nuclear matter becomes unstable,
undergoing phase transition.
With D1S the symmetric matter is unstable beyond $\rho\approx 3.4\rho_0$,
as inferred from Fig.~\ref{fig:NME_M3Yc}
and manifested in Fig.~\ref{fig:a_st}-a).
Moreover, Fig.~\ref{fig:a_st}-c) implies that
the magnetized phase emerges at moderately high $\rho$,
when we employ SLy5 or D1M.
The transition takes place at $\rho\approx 2.1\rho_0$ ($3.0\rho_0$)
in the SLy5 (D1M) result.
Similar instability occurs at $\rho\approx(1.2-3)\rho_0$
for most available parameter-sets
of the Skyrme interaction including the tensor channels,
even if deformation of the Fermi sphere is ignored~\cite{ref:CCS10}.
On the contrary, in this density region
the isotropic nuclear matter is stable against spin or isospin asymmetry
under M3Y-P6 and P7,
though M3Y-P6 gives decreasing $a_t(\rho)$ in $\rho>2.2\rho_0$,
which eventually becomes negative in $\rho>5.8\rho_0$.  

The Landau-Migdal (LM) parameters have been used
to argue global characters of excitation modes of nuclei.
Employing the analytic formulas given in Ref.~\cite{ref:Nak03},
we evaluate the LM parameters at the saturation point
for the new semi-realistic interactions, as shown in Table~\ref{tab:LM}.
See Ref.~\cite{ref:Nak03} for definition of the LM parameters.
Several LM parameters are related to the characteristic coefficients
in Table~\ref{tab:NMsat} as follows,
\begin{eqnarray}
 \frac{M^\ast_0}{M} = 1+\frac{1}{3}f_1\,,\quad
 \mathcal{K}_0 = \frac{3k_{\mathrm{F}0}^2}{M^\ast_0}(1+f_0)\,,\quad
 a_{t0} = \frac{k_{\mathrm{F}0}^2}{6M^\ast_0}(1+f'_0)\,,\nonumber\\
 a_{s0} = \frac{k_{\mathrm{F}0}^2}{6M^\ast_0}(1+g_0)\,,\quad
 a_{st0} = \frac{k_{\mathrm{F}0}^2}{6M^\ast_0}(1+g'_0)\,.
\end{eqnarray}
It has been known that $g_0$ is small
while $g'_0$ is relatively large ($\approx 1$)~\cite{ref:g'0}.
M3Y-P6 and P7 hold reasonable characters on the spin and isospin channels
as the previous parameters, 
owing significantly to $v^{(\mathrm{C})}_\mathrm{OPEP}$~\cite{ref:Nak03}.

\begin{table}
\begin{center}
\caption{Landau-Migdal parameters at the saturation point.
\label{tab:LM}}
\begin{tabular}{cr@{.}lr@{.}lr@{.}lr@{.}lr@{.}l}
\hline\hline
\hspace*{1cm} & \multicolumn{2}{c}{M3Y-P6~~} & \multicolumn{2}{c}{M3Y-P7~~} &
 \multicolumn{2}{c}{~~D1S~~~~} & \multicolumn{2}{c}{~~D1M~~~~} &
 \multicolumn{2}{c}{~~SLy5~~~} \\ \hline
$f_0$ & $-0$&$360$ & $-0$&$329$ & $-0$&$369$ & $-0$&$255$ & $-0$&$276$ \\
$f_1$ & $-1$&$211$ & $-1$&$233$ & $-0$&$909$ & $-0$&$762$ & $-0$&$909$ \\
$f_2$ & $-0$&$394$ & $-0$&$381$ & $-0$&$558$ & $-0$&$302$ & $0$&$0$ \\
$f_3$ & $-0$&$183$ & $-0$&$177$ & $-0$&$157$ & $-0$&$058$ & $0$&$0$ \\
 \hline
$f'_0$ & $0$&$544$ & $0$&$506$ & $0$&$743$ & $0$&$701$ & $0$&$815$ \\
$f'_1$ & $0$&$511$ & $0$&$571$ & $0$&$470$ & $0$&$378$ & $-0$&$387$ \\
$f'_2$ & $0$&$225$ & $0$&$234$ & $0$&$342$ & $0$&$632$ & $0$&$0$ \\
$f'_3$ & $0$&$090$ & $0$&$091$ & $0$&$100$ & $0$&$137$ & $0$&$0$ \\
 \hline
$g_0$ & $0$&$272$ & $0$&$093$ & $0$&$466$ & $-0$&$013$ & $1$&$123$ \\
$g_1$ & $0$&$231$ & $0$&$337$ & $-0$&$184$ & $-0$&$380$ & $0$&$253$ \\
$g_2$ & $0$&$163$ & $0$&$179$ & $0$&$245$ & $0$&$483$ & $0$&$0$ \\
$g_3$ & $0$&$077$ & $0$&$079$ & $0$&$091$ & $0$&$114$ & $0$&$0$ \\
 \hline
$g'_0$ & $0$&$970$ & $1$&$055$ & $0$&$631$ & $0$&$711$ & $-0$&$141$ \\
$g'_1$ & $0$&$157$ & $0$&$069$ & $0$&$610$ & $0$&$652$ & $1$&$043$ \\
$g'_2$ & $0$&$053$ & $0$&$044$ & $-0$&$038$ & $-0$&$243$ & $0$&$0$ \\
$g'_3$ & $0$&$004$ & $0$&$004$ & $-0$&$036$ & $-0$&$064$ & $0$&$0$ \\
\hline\hline
\end{tabular}
\end{center}
\end{table}

\section{Applications to finite nuclei\label{sec:FNprop}}

We next apply the new semi-realistic interactions
to finite nuclei within the MFA.
The Hamiltonian $H=H_N+V_C-H_\mathrm{c.m.}$ is used,
with $H_N$ given in Eq.~(\ref{eq:H_N}).
$V_C$ and $H_\mathrm{c.m.}$ represent the Coulomb interaction
and the center-of-mass (c.m.) Hamiltonian.
We have made no additional approximation on $H$,
by handling the exchange term of $V_C$
and the two-body terms of $H_\mathrm{c.m.}$ explicitly.
Effects of $V_C$ on the proton pairing,
which have recently been recognized to be sizable~\cite{ref:Coul-pair},
are explicitly included in the Hartree-Fock-Bogolyubov (HFB) calculations
as well.

The algorithm based on the Gaussian expansion method
(GEM)~\cite{ref:NS02,ref:Nak06} is applied
for all the numerical calculations of finite nuclei in this article.
In this method we employ the s.p. basis-functions of
\begin{eqnarray} \varphi_{\nu\ell jm}(\mathbf{r})
= R_{\nu\ell j}(r)[Y^{(\ell)}(\hat{\mathbf{r}})\chi_\sigma]^{(j)}_m\,;
\quad
R_{\nu\ell j}(r) = \mathcal{N}_{\nu\ell j}\,
r^\ell \exp(-\nu r^2)\,,
\label{eq:basis} \end{eqnarray}
apart from the isospin index,
where $Y^{(\ell)}(\hat{\mathbf{r}})$ is the spherical harmonics
and $\chi_\sigma$ the spin wave function.
For the range parameter $\nu$,
which is generally a complex number ($\nu=\nu_\mathrm{r}+i\nu_\mathrm{i}$),
we adopt the following values~\cite{ref:Nak08}:
\begin{equation}
\nu_\mathrm{r}=\nu_0\,b^{-2n}\,,\quad
\left\{\begin{array}{ll}\nu_\mathrm{i}=0 & (n=0,1,\cdots,5)\\
{\displaystyle\frac{\nu_\mathrm{i}}{\nu_\mathrm{r}}
=\pm\frac{\pi}{2}} & (n=0,1,2)\end{array}\right.\,,
 \label{eq:basis-param}
\end{equation}
with $\nu_0=(2.40\,\mathrm{fm})^{-2}$ and $b=1.25$,
resulting in $12$ bases for each $(\ell,j)$.
In the HFB calculations the s.p. space is truncated
as $\ell\leq 8$ (for the $Z=82$ and $N=126$ nuclei)
or $\ell\leq 7$ (for the lighter ones).
As shown in Ref.~\cite{ref:Nak08},
the above set of the GEM bases can cover
wide range of nuclear mass with good precision.

In the following we shall show results
of the spherical HF and HFB calculations
using the new M3Y-P6 and P7 interactions,
in comparison with those using the Gogny D1S and D1M interactions.
Although there have been more advanced calculations
using the Gogny interactions, \textit{e.g.} the calculations
based on the generator-coordinate method~\cite{ref:Ber07},
comparison is made only at the MFA level in this paper,
leaving extensive applications of the M3Y-type interactions as a future work.

\subsection{Doubly magic nuclei\label{subsec:DM}}

The spherical HF approach is rationally expected to be a good approximation
for the ground states of the doubly magic nuclei.
We present the binding energies and the rms matter radii
of several doubly magic nuclei in Table~\ref{tab:DMprop}.
The spherical HF results using the new semi-realistic interactions
are compared with those using D1S and D1M,
as well as with the experimental data.
Influence of the c.m. motion on the matter radii
is subtracted in a similar manner to the c.m. energies~\cite{ref:Nak03}.
The binding energies of these nuclei by the M3Y-P6 and P7 interactions
are in agreement with the measured values within $5\,\mathrm{MeV}$ accuracy,
except $^{40}$Ca.
This accuracy is comparable to those of D1S and D1M.
In $^{40}$Ca, influence of octupole correlations might be strong,
as suggested by the low $3^-_1$ energy in measurements~\cite{ref:TI}
and mentioned in Ref.~\cite{ref:Nak08b}.
For this reason we have not taken this discrepancy seriously
at the present stage,
while future study is needed on this problem.
The rms matter radii of these nuclei calculated from M3Y-P6 and P7
are also in fair agreement with the data.
We point out that D1S has not predicted accurate energy of $^{100}$Sn,
and that D1M systematically gives smaller radii than the measured ones.

\begin{table}
\begin{center}
\caption{Binding energies ($-E$) and rms matter radii
 ($\sqrt{\langle r^2\rangle}$) of several doubly magic nuclei.
 Experimental data are taken
 from Refs.~\protect\cite{ref:mass,ref:O16-rad,ref:O24-rad,ref:rad}.
\label{tab:DMprop}}
\begin{tabular}{cccrrrrrr}
\hline\hline
&&&~~~Exp.~~&~~M3Y-P6~&~~M3Y-P7~&~~~~~D1S~~&~~~~~D1M~~\\ \hline
$^{16}$O & $-E$ &(MeV)&
 $127.6$ & $126.3$ & $125.9$ & $129.5$ & $128.2$ \\
& $\sqrt{\langle r^2\rangle}$ &(fm)&
 $2.61$ & $2.59$ & $2.57$ & $2.61$ & $2.57$ \\
$^{24}$O & $-E$ &(MeV)&
 $168.5$ & $166.2$ & $167.4$ & $168.6$ & $167.3$ \\
& $\sqrt{\langle r^2\rangle}$ &(fm)&
 $3.19$ & $3.05$ & $3.03$ & $3.01$ & $2.98$ \\
$^{40}$Ca & $-E$ &(MeV)&
 $342.1$ & $335.9$ & $334.3$ & $344.6$ & $342.2$ \\
& $\sqrt{\langle r^2\rangle}$ &(fm)&
 $3.47$ & $3.37$ & $3.35$ & $3.37$ & $3.33$ \\
$^{48}$Ca & $-E$ &(MeV)&
 $416.0$ & $413.8$ & $414.9$ & $416.8$ & $414.6$ \\
& $\sqrt{\langle r^2\rangle}$ &(fm)&
 $3.57$ & $3.51$ & $3.49$ & $3.51$ & $3.48$ \\
$^{90}$Zr & $-E$ &(MeV)&
 $783.9$ & $781.1$ & $780.8$ & $785.9$ & $782.1$ \\
& $\sqrt{\langle r^2\rangle}$ &(fm)&
 $4.32$ & $4.23$ & $4.22$ & $4.24$ & $4.20$ \\
$^{100}$Sn & $-E$ &(MeV)&
 $824.8$ & $822.5$ & $822.8$ & $831.6$ & $824.9$ \\
& $\sqrt{\langle r^2\rangle}$ &(fm)&
 --- & $4.36$ & $4.34$ & $4.36$ & $4.32$ \\
$^{132}$Sn & $-E$ &(MeV)&
 $1102.9$ & $1097.8$ & $1100.8$ & $1104.1$ & $1104.5$ \\
& $\sqrt{\langle r^2\rangle}$ &(fm)&
 --- & $4.78$ & $4.77$ & $4.77$ & $4.72$ \\
$^{208}$Pb & $-E$ &(MeV)&
 $1636.4$ & $1634.5$ & $1635.5$ & $1639.0$ & $1638.9$ \\
& $\sqrt{\langle r^2\rangle}$ &(fm)&
 $5.49$ & $5.53$ & $5.51$ & $5.51$ & $5.47$ \\
\hline\hline
\end{tabular}
\end{center}
\end{table}

\begin{figure}
\includegraphics[scale=1.0]{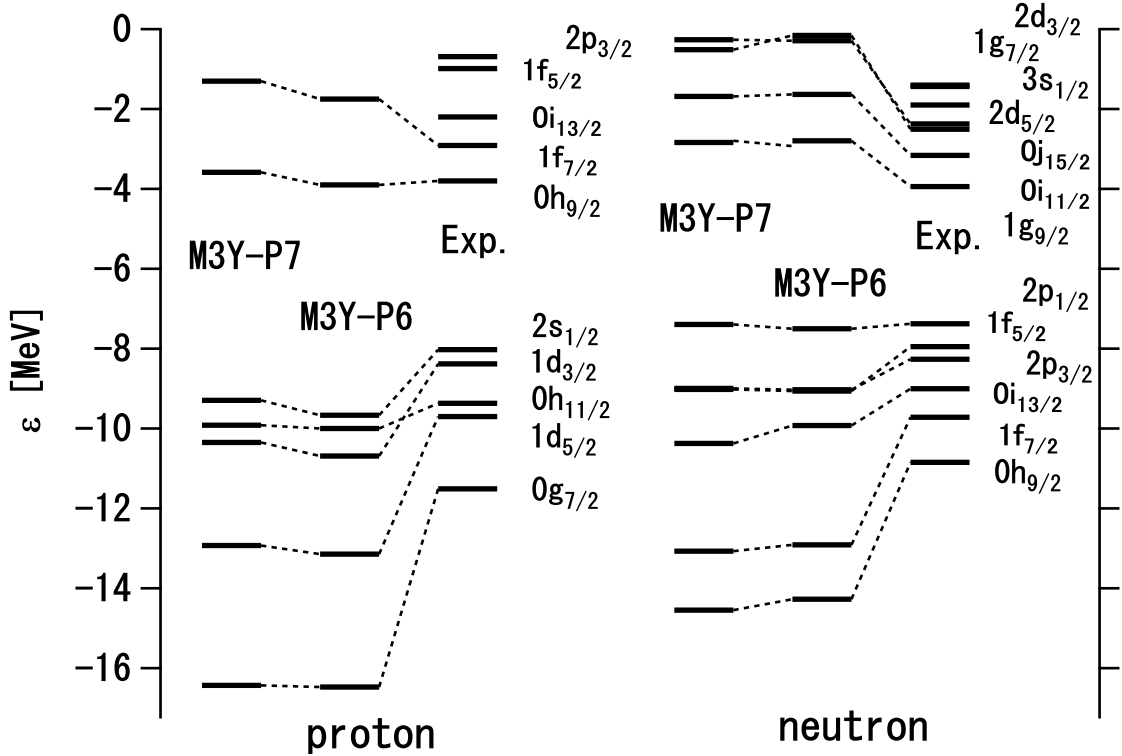}
\caption{Single-particle energies for $^{208}$Pb.
Experimental values are extracted
from Refs.~\protect\cite{ref:mass,ref:TI}.
\label{fig:Pb_spe}}
\end{figure}

The non-central channels of the effective interaction,
$v^{(\mathrm{LS})}$ and $v^{(\mathrm{TN})}$,
are responsible for the $\ell s$ splitting of the s.p. levels
and its nucleus-dependence.
We display the s.p. levels of $^{208}$Pb calculated in the HF approximation,
comparing to the observed levels in Fig.~\ref{fig:Pb_spe}.
The experimental s.p. levels are taken from the lowest states
having specific spin-parity in the $A=207$ or $209$ nuclei.
Because of the fragmentation via the coupling
to the many-particle-many-hole configurations,
these observed states do not straightforwardly correspond
to the s.p. levels in the MFA.
In M3Y-P6 and P7 the non-central channels are not changed from M3Y-P0
except the overall enhancement factor to $v^{(\mathrm{LS})}$.
This factor is determined so that the level ordering
should not differ seriously from the observed one
around $^{208}$Pb.
While appropriateness of the enhancement factor to $v^{(\mathrm{LS})}$
should further be investigated in future studies,
it is a simple and useful cure to the $\ell s$ splitting. 

It has been recognized that the neutron-skin thickness in $^{208}$Pb
is connected to the $\rho$ dependence of the symmetry energy~\cite{ref:CR09},
particularly the $\mathcal{L}_{t0}$ parameter.
Table~\ref{tab:n-skin_Pb208} presents
difference between proton and neutron rms radii
$\sqrt{\langle r^2\rangle_n} - \sqrt{\langle r^2\rangle_p}$ in $^{208}$Pb.
The values of M3Y-P6 and P7 are in good agreement
with the recent experimental value
drawn from the $E1$ strengths~\cite{ref:Tam11}.

\begin{table}
\begin{center}
\caption{Difference between proton and neutron rms radii
$\sqrt{\langle r^2\rangle_n} - \sqrt{\langle r^2\rangle_p}$
in $^{208}$Pb ($\mathrm{fm}$)
by the HF calculations with several interactions.
\label{tab:n-skin_Pb208}}
\begin{tabular}{cccc}
\hline\hline
~M3Y-P6~&~M3Y-P7~&~~~D1S~~~&~~~D1M~~~\\ \hline
 $0.158$ & $0.161$ & $0.136$ & $0.112$ \\
\hline\hline
\end{tabular}
\end{center}
\end{table}

\subsection{Proton- or neutron-magic nuclei\label{subsec:SM}}

The spherical HFB approach provides us with a reasonable approximation
for the nuclei in which $Z$ or $N$ is a magic number.
The odd-$Z$ or $N$ nuclei can be handled
in the equal-filling approximation~\cite{ref:EFA,ref:EFAb}.
In fixing the new parameter-sets,
we have taken into account the pairing properties
by fitting to the data on the even-odd mass differences
in the $Z=50$, $N\sim 70$ and the $N=82$, $Z\sim 60$ nuclei~\cite{footnote}.

\begin{figure}
\includegraphics[scale=0.6]{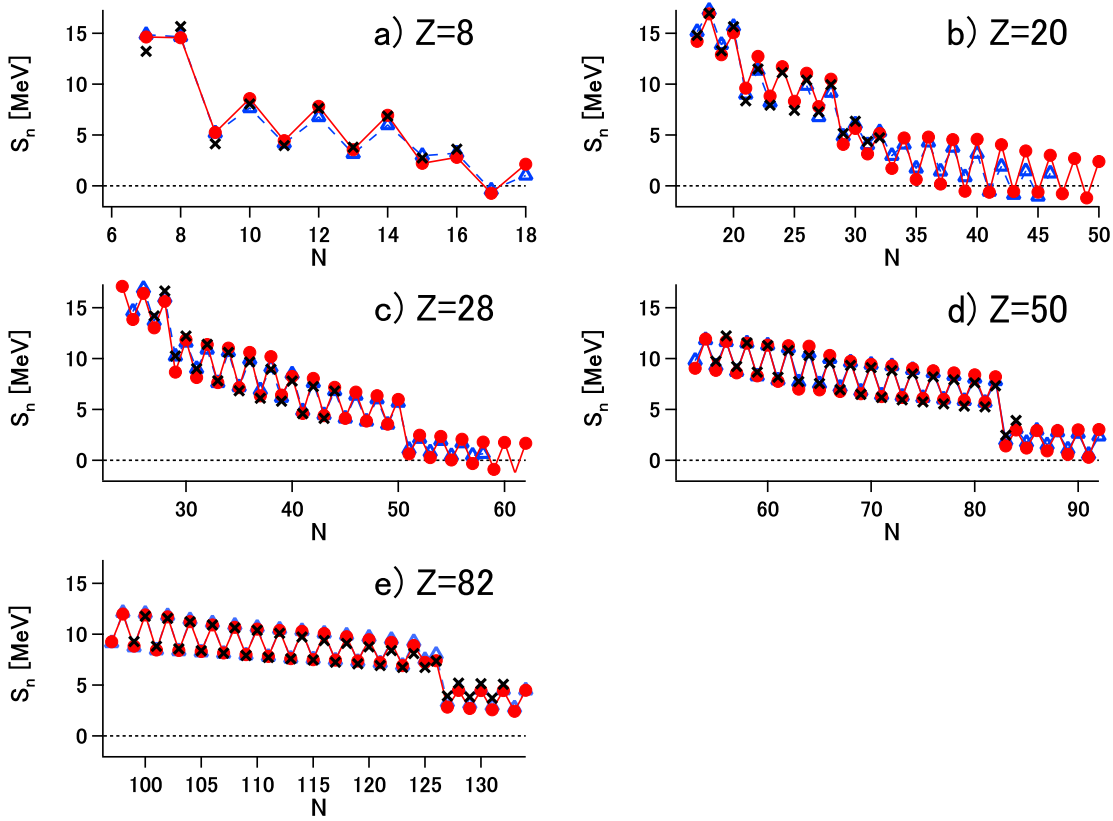}
\caption{(Color online) Neutron separation energies
for a) $Z=8$, b) $Z=20$, c) $Z=28$, d) $Z=50$ and e) $Z=82$ nuclei,
calculated with M3Y-P7 (red circles) and D1M (blue open triangles).
Lines are drawn to guide eyes.
Experimental values are taken from Ref.~\protect\cite{ref:mass}
and presented by the crosses.
\label{fig:Sn}}
\end{figure}

\begin{figure}
\includegraphics[scale=0.6]{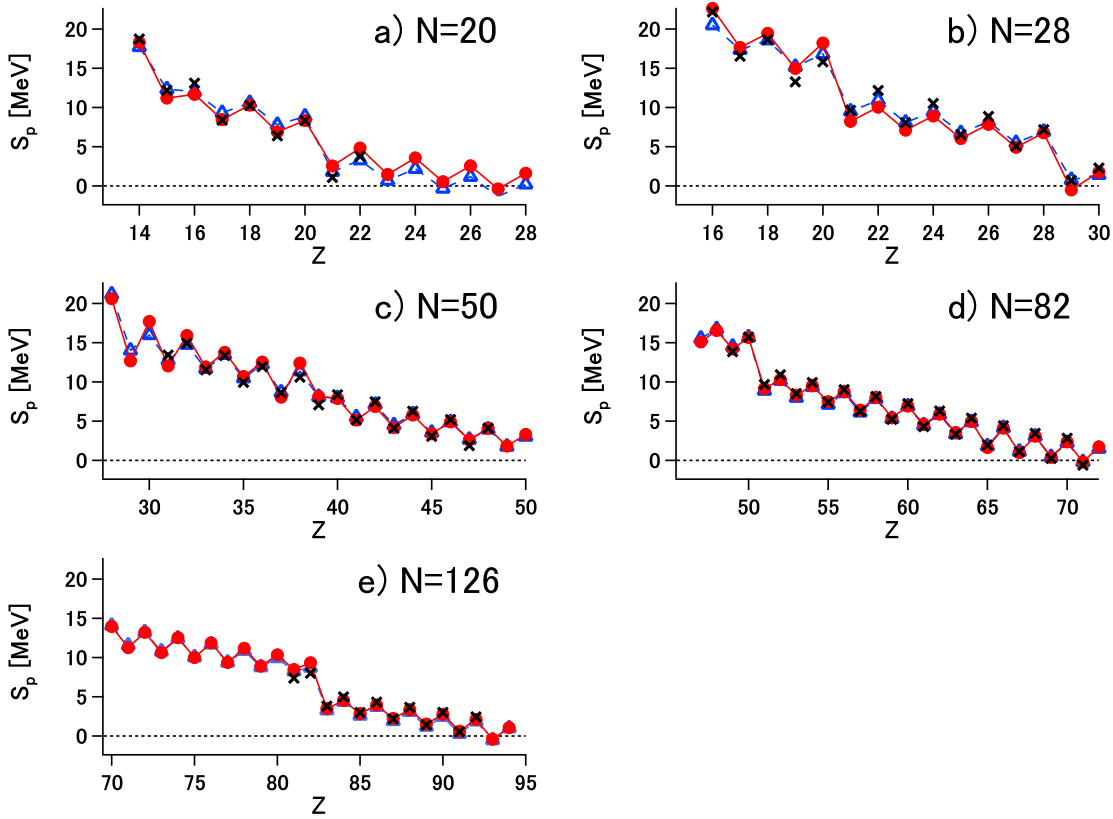}
\caption{(Color online) Proton separation energies
for a) $N=20$, b) $N=28$, c) $N=50$, d) $N=82$ and e) $N=126$ nuclei.
See Fig.~\protect\ref{fig:Sn} for the conventions.
\label{fig:Sp}}
\end{figure}

In Fig.~\ref{fig:Sn} (Fig.~\ref{fig:Sp}),
the neutron (proton) separation energies $S_n$ ($S_p$) are plotted
for the $Z=\mathrm{magic}$ ($N=\mathrm{magic}$) nuclei.
The $S_n$ and $S_p$ values of M3Y-P6,
which are always close to those of M3Y-P7, are not presented.
Similarly, being close to the D1M ones, the D1S results are not displayed.
Notice that the even-odd mass difference
is proportional to the difference of the separation energies
between the adjacent nuclei,
while the two-neutron (two-proton) separation energy
is the sum of $S_n$'s ($S_p$'s) of the two neighboring nuclei.
Although there are certain discrepancies if we look into their details,
M3Y-P6 and P7 give separation energies
in agreement with the measured ones
with the accuracy similar to D1S and D1M.

It has been suggested that isotope shifts of the Pb nuclei
may be relevant to the $\ell s$ potential~\cite{ref:SLKP95},
which is primarily determined by $v^{(\mathrm{LS})}$.
In $\delta\sqrt{\langle r^2\rangle_p}(\mbox{$^A$Pb})
=\sqrt{\langle r^2\rangle_p}(\mbox{$^A$Pb})
-\sqrt{\langle r^2\rangle_p}(\mbox{$^{208}$Pb})$,
a kink has been observed at $^{208}$Pb.
The zero-range LS force contained in the Gogny
as well as the original Skyrme interactions
operates only on the $T=1$ two-nucleon states.
It was argued that this isospin character of the LS force
could be insufficient to describe
the rapid rise of $\delta\sqrt{\langle r^2\rangle_p}$ in $N>126$.
Having $v^{(\mathrm{LS})}$ with finite ranges
that acts also on the $T=0$ channel,
it may be interesting whether the M3Y-type interactions
reproduce the kink of $\delta\sqrt{\langle r^2\rangle_p}$.

\begin{figure}
\includegraphics[scale=0.8]{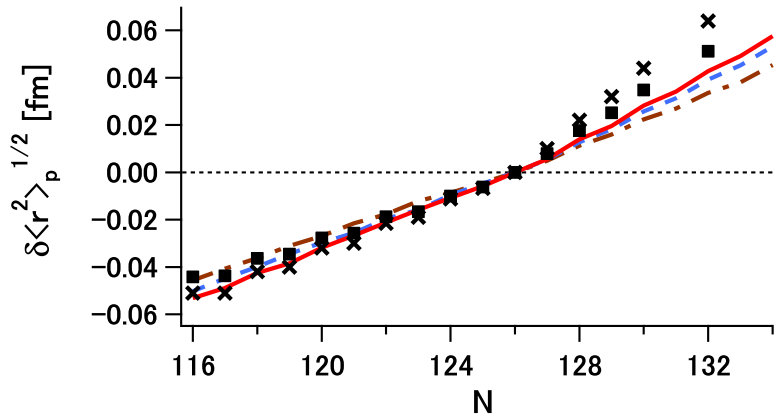}
\caption{(Color online) Isotope shifts of the Pb nuclei
$\delta\sqrt{\langle r^2\rangle_p}(\mbox{$^A$Pb})$,
obtained from the HFB calculations with M3Y-P7 (red solid line),
D1M (blue dashed line) and D1S (brown dot-dashed line).
Experimental data are taken from Refs.~\cite{ref:AHS87}
and \cite{ref:Ang04},
which are represented by squares and crosses, respectively.
\label{fig:Pb_drp}}
\end{figure}

In Fig.~\ref{fig:Pb_drp},
we depict $\delta\sqrt{\langle r^2\rangle_p}(\mbox{$^A$Pb})$
obtained by the spherical HFB calculations with M3Y-P7, D1S and D1M,
in comparison with experimental data~\cite{ref:AHS87,ref:Ang04}.
The results of M3Y-P6 are almost indistinguishable
from those of D1M in $N<126$ and from those of M3Y-P7 in $N>126$.
The D1S interaction does not give clear bending at $^{208}$Pb.
On the contrary, although its LS force holds the zero-range form,
D1M provides a visible kink at $^{208}$Pb.
M3Y-P7, in which $v^{(\mathrm{LS})}$ has finite ranges,
further improves $\delta\sqrt{\langle r^2\rangle_p}$.
We have confirmed that this tendency well correlates
to the occupation probability on $n0i_{11/2}$,
as pointed out in Ref.~\cite{ref:RF95}.
It seems reasonable to consider
that the s.p. energy difference $\varepsilon(n0i_{11/2})-\varepsilon(n1g_{9/2})$
is responsible for the interaction-dependence
of $\delta\sqrt{\langle r^2\rangle_p}$ in $N>126$.
However, whereas isospin character of $v^{(\mathrm{LS})}$
plays a certain role in the s.p. energy difference,
it is not yet obvious whether and how interplay of other channels,
\textit{e.g.} $v^{(\mathrm{TN})}$ and the pairing,
contributes to $\delta\sqrt{\langle r^2\rangle_p}$.

In earlier studies~\cite{ref:SLKP95,ref:RF95},
the MFA results were compared with the data given in,
\textit{e.g.}, Ref.~\cite{ref:AHS87}.
The kink at $^{208}$Pb becomes stronger in the new data~\cite{ref:Ang04}
because of larger values of $\delta\sqrt{\langle r^2\rangle_p}$ in $N>126$,
which has not been reproduced within the MFA, to the author's best knowledge.

\subsection{Shell structure and tensor force\label{subsec:shell}}

It has been established that the shell structure depends on
the effective $NN$ interaction of the MFA.
In particular, the effects of the tensor force are significant,
as clarified in the proton- or neutron-magic nuclei~\cite{ref:Nak08b,ref:Vtn}.
In some nuclei the spin-isospin channels of the central force
could also have appreciable effects~\cite{ref:Nak08b,ref:Vst}.
Having realistic $v^{(\mathrm{TN})}$ and $v^{(\mathrm{C})}_\mathrm{OPEP}$,
M3Y-P6 and P7 are suitable to investigating those effects
as the previous interaction M3Y-P5~\cite{ref:Nak08b,ref:Nak10}.
This is a clear advantage of these semi-realistic interactions
over the phenomenological interactions such as D1S and D1M.
Although the results of M3Y-P6 and P7 on the shell structure
are not essentially different from those of the previous parameter-set
shown in Refs.~\cite{ref:Nak08b,ref:Nak10,ref:Nak10b},
we briefly mention several notable points.

The $N$-dependence of the differences of the observed s.p. energies
$\varepsilon(p0h_{11/2})-\varepsilon(p1d_{5/2})$ and
$\varepsilon(p0g_{7/2})-\varepsilon(p1d_{5/2})$
in the Sn isotopes are simultaneously reproduced,
with $v^{(\mathrm{TN})}$ being crucial to the former
(see Fig.~14 and related arguments in Ref.~\cite{ref:Nak08b}).
In the $N=32$ isotones $v^{(\mathrm{TN})}$ and $v^{(\mathrm{C})}_\mathrm{OPEP}$
give rise to significant $Z$-dependence
of $\varepsilon(n0f_{5/2})-\varepsilon(n1p_{3/2})$
(see Fig.~13 of Ref.~\cite{ref:Nak08b}
and Fig.~3 of Ref.~\cite{ref:Nak10b}),
possibly accounting for the new magic number $N=32$
around $^{52}$Ca~\cite{ref:N32,ref:Ca52_Ex2}.
However, despite the presence of the realistic tensor force,
the semi-realistic interactions do not predict closure of $N=34$ at $^{54}$Ca
(see Fig.~3 of Ref.~\cite{ref:Nak10}),
in contrast to the shell model prediction in Ref.~\cite{ref:GXPF1A}.
While semi-magic nature of $N=40$ is indicated at $^{68}$Ni
with the semi-realistic interactions as with the Gogny interactions
(see Fig.~4 of Ref.~\cite{ref:Nak10}),
which seems consistent with experiments~\cite{ref:Ni68},
it is likely for the $N=40$ magic nature to be broken at $^{60}$Ca
because of $v^{(\mathrm{TN})}$ (see Fig.~3 of Ref.~\cite{ref:Nak10}
and Fig.~4 of Ref.~\cite{ref:Nak10b}).
It is of interest to investigate magicity of $N=40$ toward $^{60}$Ca
experimentally,
which could further clarify role of $v^{(\mathrm{TN})}$.
The doubly magic nature of $^{78}$Ni is predicted to hold
even with $v^{(\mathrm{TN})}$,
giving $E_x(2^+_1)=3.0-3.5\,\mathrm{MeV}$
and $B(E2;2^+_1\to 0^+_1)\approx 85\,e^2\mathrm{fm}^4$ in the HF+RPA.
Experimental data on $2^+_1$ of $^{78}$Ni are awaited.

Location of the neutron drip line for the Ca and Ni nuclei
could be investigated by the new experimental facilities~\cite{ref:RIB07}
and has been argued in Ref.~\cite{ref:Nak10b}.
We tabulate the neutron drip line predicted by the spherical HFB calculations
with the M3Y-type and the Gogny interactions, in Table~\ref{tab:n-drip}.

\begin{table}
\begin{center}
\caption{Neutron numbers of the heaviest bound Ca and Ni nuclei
 predicted by the spherical HFB calculations
 with several interactions.
\label{tab:n-drip}}
\begin{tabular}{ccccc}
\hline\hline
Isotope &~M3Y-P6~&~M3Y-P7~&~~~D1S~~~&~~~D1M~~~\\ \hline
Ca & $50$ & $50$ & $44$ & $46$ \\
Ni & $60$ & $62$ & $58$ & $58$ \\
\hline\hline
\end{tabular}
\end{center}
\end{table}

\section{Summary and outlook\label{sec:summary}}

We have developed new parameter-sets
of the semi-realistic effective interactions
to describe low energy phenomena of nuclei.
They are obtained by phenomenologically modifying several parameters
in the M3Y-Paris interaction,
while the tensor force and the OPEP part in the central force
are not changed,
as before.
Unlike the previous parameters,
the new sets M3Y-P6 and P7 are adjusted
also to the microscopic (FP and APR) results of the neutron-matter energies
and to the binding energy of $^{100}$Sn.
We therefore attain improvement on the symmetry energy,
up to its density-dependence.
In contrast to instability of the spin-saturated symmetric nuclear matter
in the SLy5, D1S and D1M results,
neither of M3Y-P6 nor P7 predicts such phase transition
in the density range of $\rho\lesssim 6\rho_0$.

The new parameter-sets M3Y-P6 and P7 have been applied
to the doubly magic nuclei in the spherical HF calculations,
and to the proton- or neutron-magic nuclei in the spherical HFB calculations.
Fair agreement with experimental data has been demonstrated
for the binding energies of the doubly magic nuclei
and for the nucleon separation energies
of the proton- or neutron-magic nuclei.
Owing to the realistic tensor force and the OPEP central force,
the $Z$- or $N$-dependence of the shell structure
is well described with M3Y-P6 and P7,
as with the previous set M3Y-P5.
The isotope shifts of the Pb nuclei have also been argued.

Future study includes application of the semi-realistic interactions
to the excitations in the RPA,
as well as to deformed nuclei.
Moreover, extensive applications to nuclear reactions
and to the neutron stars may be within the reach,
which give further test of the effective interactions
and a step toward unified description of nuclear structure,
reactions and neutron stars.

\begin{acknowledgments}
This work is financially supported
as Grant-in-Aid for Scientific Research (C), No.~22540266,
by Japan Society for the Promotion of Science,
and as Grant-in-Aid for Scientific Research on Innovative Areas,
No.~24105008, by The Ministry of Education, Culture, Sports, Science
and Technology, Japan.
Numerical calculations are performed on HITAC SR16000s
at Institute of Media and Information Technology in Chiba University,
Yukawa Institute for Theoretical Physics in Kyoto University,
Research Institute for Information Technology in Kyushu University,
Information Technology Center in University of Tokyo,
and Information Initiative Center in Hokkaido University.
\end{acknowledgments}


\begin{thebibliography}{99}
\bibitem{ref:micro} S.C. Pieper, K. Varga and R.B. Wiringa,
 Phys. Rev. C \textbf{66}, 044310 (2002).
\bibitem{ref:NCSM} P. Navrat\'{i}l and B.R. Barrett,
 Phys. Rev. C \textbf{57}, 3119 (1998);
P. Navrat\'{i}l, J.P. Vary and B.R. Barrett,
 Phys. Rev. C \textbf{62}, 054311 (2000);
R. Roth, J. Langhammer, A. Calci, S. Binder and P. Navrat\'{i}l,
 Phys. Rev. Lett. \textbf{107}, 072501 (2011).
\bibitem{ref:HPDH08} G. Hagen, T. Papenbrock, D.J. Dean and
 M. Hjorth-Jensen, Phys. Rev. Lett. \textbf{101}, 092502 (2008);
G. Hagen, M. Hjorth-Jensen, G.R. Jansen, R. Machleidt and T. Papenbrock,
 Phys. Rev. Lett. \textbf{109}, 032502 (2012).
\bibitem{ref:Nak03} H. Nakada, Phys. Rev. C \textbf{68}, 014316 (2003).
\bibitem{ref:Nak08b} H. Nakada, Phys. Rev. C \textbf{78}, 054301 (2008);
 \textit{ibid.} \textbf{82}, 029902(E) (2010).
\bibitem{ref:M3Y} G. Bertsch, J. Borysowicz, H. McManus and W.G. Love,
 Nucl. Phys. A \textbf{284}, 399 (1977).
\bibitem{ref:M3Y-P} N. Anantaraman, H. Toki and G.F. Bertsch,
 Nucl. Phys. A \textbf{398}, 269 (1983).
\bibitem{ref:NS02} H. Nakada and M. Sato,
 Nucl. Phys. A \textbf{699}, 511 (2002);
 \textit{ibid.} \textbf{714}, 696 (2003).
\bibitem{ref:Nak06} H. Nakada, Nucl. Phys. A \textbf{764}, 117 (2006);
 \textit{ibid.} \textbf{801}, 169 (2008).
\bibitem{ref:Nak08} H. Nakada, Nucl. Phys. A \textbf{808}, 47 (2008).
\bibitem{ref:NMYM09} H. Nakada, K. Mizuyama, M. Yamagami and M. Matsuo,
 Nucl. Phys. A \textbf{828}, 283 (2009).
\bibitem{ref:Nak04} H. Nakada, \textit{Proceedings of
 the International Symposium ``A New Era of Nuclear Structure Physics},
 edited by Y. Suzuki, M. Matsuo, S. Ohya and T. Ohtsubo, p.~184
 (World Scientific, Singapore, 2004).
\bibitem{ref:Nak09} H. Nakada, Eur. Phys. J. A \textbf{42}, 565 (2009).
\bibitem{ref:Nak10b} H. Nakada, Phys. Rev. C \textbf{81}, 051302(R) (2010).
\bibitem{ref:Shi08} T. Shizuma \textit{et al.},
 Phys. Rev. C \textbf{78}, 061303(R) (2008).
\bibitem{ref:Vst} T. Otsuka \textit{et al.},
 Phys. Rev. Lett. \textbf{87}, 082502 (2001).
\bibitem{ref:Vtn} T. Otsuka, T. Suzuki, R. Fujimoto, H. Grawe
 and Y. Akaishi, Phys. Rev. Lett. {\bf 95}, 232502 (2005).
\bibitem{ref:Nak10} H. Nakada, Phys. Rev. C \textbf{81}, 027301 (2010);
 \textit{ibid.} \textbf{82}, 029903(E) (2010).
\bibitem{ref:Khoa-pv} D.T. Khoa, private communication.
\bibitem{ref:LTKM11} D.T. Loan, N.H. Tan, D.T. Khoa
 and J. Margueron, Phys. Rev. C \textbf{83}, 065809 (2011).
\bibitem{ref:LP07} J.M. Lattimer and M. Prakash,
 Phys. Rep. \textbf{442}, 109 (2007).
\bibitem{ref:Ono03} A. Ono, P. Danielewicz, W.A. Friedman, W.G. Lynch
 and M.B. Tsang, Phys. Rev. C \textbf{68}, 051601(R) (2003).
\bibitem{ref:PDR-L} A. Klimkiewicz \textit{et al.},
 Phys. Rev. C \textbf{76}, 051603(R) (2007);
A. Carbone \textit{et al.}, Phys. Rev. C \textbf{81}, 041301(R) (2010).
\bibitem{ref:TKG09} H.S. Than, D.T. Khoa and N.V. Giai,
 Phys. Rev. C \textbf{80}, 064312 (2009).
\bibitem{ref:APR98} A. Akmal, V.R. Pandharipande and D.G. Ravenhall,
 Phys. Rev. C \textbf{58}, 1804 (1998).
\bibitem{ref:FP81} B. Friedman and V.R. Pandharipande,
 Nucl. Phys. A \textbf{361}, 502 (1981).
\bibitem{ref:PDG06} ParticleDataGroup, J. Phys. G \textbf{33}, 1 (2006).
\bibitem{ref:LS} K. Suzuki, R. Okamoto and H. Kumagai,
 Phys. Rev. C \textbf{36}, 804 (1987);
S.C. Pieper and V.R. Pandharipande,
 Phys. Rev. Lett. \textbf{70}, 2541 (1993).
\bibitem{ref:SLy} E. Chabanat, P. Bonche, P. Haensel, J. Meyer
 and R. Schaeffer, Nucl. Phys. A \textbf{635}, 231 (1998).
\bibitem{ref:D1S} J.F. Berger, M. Girod and D. Gogny,
 Comp. Phys. Comm. \textbf{63}, 365 (1991).
\bibitem{ref:D1M} S. Goriely, S. Hilaire, M. Girod and S. P\`{e}ru,
 Phys. Rev. Lett. \textbf{102}, 242501 (2009).
\bibitem{ref:WFF88} R.B. Wiringa, V. Fiks and A. Fabrocini,
 Phys. Rev. C \textbf{38}, 1010 (1988).
\bibitem{ref:SKC06} S. Shlomo, M. Kolomietz and G. Col\`{o},
 Eur. Phys. J. A \textbf{30}, 23 (2006).
\bibitem{ref:Dan03} P. Danielewicz, Nucl. Phys. A \textbf{727}, 233 (2003).
\bibitem{ref:UNDEF1} M. Kortelainen \textit{et al.},
 Phys. Rev. C \textbf{85}, 024304 (2012).
\bibitem{ref:Mahaux} C. Mahaux, P.F. Bortignon, R.A. Broglia
and C.H. Dasso, Phys. Rep. \textbf{120}, 1 (1985).
\bibitem{ref:CCS10} L.-G. Cao, G. Col\`{o} and H. Sagawa,
 Phys. Rev. C \textbf{81}, 044302 (2010).
\bibitem{ref:g'0} C. Gaarde \textit{et al.},
 Nucl. Phys. A \textbf{369}, 258 (1981);
T. Suzuki, Nucl. Phys. A \textbf{379}, 110 (1982);
G. Bertsch, D. Cha and H. Toki, Phys. Rev. C \textbf{24}, 533 (1981);
T. Suzuki and H. Sakai, Phys. Lett. B \textbf{455}, 25 (1999).
\bibitem{ref:Coul-pair} M. Anguiano, J.L. Egido and L.M. Robledo,
 Nucl. Phys. A \textbf{683}, 227 (2001);
T. Lesinski, T. Duguet, K. Bennaceur and J. Meyer,
 Eur. Phys. J. A \textbf{40}, 121 (2009);
H. Nakada and M. Yamagami,
 Phys. Rev. C \textbf{83}, 031302(R) (2011).
\bibitem{ref:Ber07} G.F. Bertsch \textit{et al.},
 Phys. Rev. Lett. \textbf{99}, 032502 (2007).
\bibitem{ref:mass} G. Audi and A.H. Wapstra,
 Nucl. Phys. A \textbf{595}, 409 (1995);
G. Audi, A.H. Wapstra and C. Thibault,
 Nucl. Phys. A \textbf{729}, 337 (2003).
\bibitem{ref:O16-rad} D.T. Khoa, H.S. Than and M. Grasso,
 Nucl. Phys. A \textbf{722}, 92c (2003).
\bibitem{ref:O24-rad} A. Ozawa \textit{et al.},
 Nucl. Phys. A \textbf{691} (2001) 599.
\bibitem{ref:rad} G.D. Alkhazov, S.L. Belostotsky and A.A. Vorobyov,
 Phys. Rep. \textbf{42}, 89 (1978).
\bibitem{ref:TI} R.B. Firestone \textit{et al.},
\textit{Table of Isotopes}, 8th edition
(John Wiley \& Sons, New York, 1996).
\bibitem{ref:CR09} M. Centelles, X. Roca-Maza, X. Vi\~{n}as and M. Warda,
 Phys. Rev. Lett. \textbf{102}, 122502 (2009).
\bibitem{ref:Tam11} A. Tamii \textit{et al.},
 Phys. Rev. Lett. \textbf{107}, 062502 (2011).
\bibitem{ref:EFA} S. Perez-Martin and L.M. Robledo,
 Phys. Rev. C \textbf{78}, 014304 (2008).
\bibitem{ref:EFAb} N. Schunck \textit{et al.},
 Phys. Rev. C \textbf{81}, 024316 (2010).
\bibitem{footnote} Although the set M3Y-P5$'$ was claimed to be fitted
to the even-odd mass difference in the $Z=50$ nuclei
by the fully self-consistent HFB calculations~\cite{ref:Nak10},
inconsistency had remained in evaluation of contribution of $v^{(\mathrm{DD})}$
to pairing energy.
\bibitem{ref:SLKP95} M.M. Sharma, G. Lalazissis, J. K\"{o}nig and P. Ring,
 Phys. Rev. Lett. \textbf{74}, 3744 (1995).
\bibitem{ref:AHS87} P. Aufmuth, K. Heilig and A. Steudel,
 At. Data Nucl. Data Tables \textbf{37}, 455 (1987).
\bibitem{ref:Ang04} I. Angeli,
 At. Data Nucl. Data Tables \textbf{87}, 185 (2004).
\bibitem{ref:RF95} P.-G. Reinhard and H. Flocard,
 Nucl. Phys. A \textbf{584}, 467 (1995).
\bibitem{ref:N32} R. Kanungo, I. Tanihata and A. Ozawa,
 Phys. Lett. B \textbf{528} (2002) 58.
\bibitem{ref:Ca52_Ex2} J.I. Prisciandaro \textit{et al.},
 Phys. Lett. B \textbf{510}, 17 (2001).
\bibitem{ref:GXPF1A} M. Honma, T. Otsuka. B.A. Brown and T. Mizusaki,
 Eur. Phys. J. A \textbf{25}, s01, 499 (2005).
\bibitem{ref:Ni68} R. Broda \textit{et al.},
 Phys. Rev. Lett. \textbf{74}, 868 (1995).
\bibitem{ref:RIB07} T. Aumann, Prog. Part. Nucl. Phys.
 \textbf{59}, 3 (2007);
S. Gales, Prog. Part. Nucl. Phys. \textbf{59}, 22 (2007);
T. Motobayashi, Prog. Part. Nucl. Phys. \textbf{59}, 32 (2007).

\end{thebibliography}

\end{document}